\documentclass[12pt, longbibliography]{revtex4-1}

\usepackage{graphicx}% Include figure files
\usepackage{dcolumn}% Align table columns on decimal point
\usepackage{bm}% bold math

\newcommand{\dograph}[2]{
	\begin{figure}
	\includegraphics[width=6in]{#1}
	\caption{\label{fig:#1}#2}
	\end{figure}
}

\begin{document}

\title{Evidence for an anomalous current-phase relation in topological insulator Josephson junctions}
\author{C.~Kurter$^{1*}$, A.D.K.~Finck$^{1*}$, Y.S.~Hor$^2$, D.J.~Van Harlingen$^{1\dag}$}
\affiliation{$^1$Department of Physics and Materials Research Laboratory, University of Illinois at Urbana-Champaign, Urbana, Illinois 61801
\\
$^2$Department of Physics, Missouri University of Science and Technology, Rolla, MO 65409 \\ $^{*}$ These authors contributed equally to the work.\\$^{\dag}$ Corresponding author: dvh@illinois.edu}
\date{\today}

\begin{abstract}
{\bf
Josephson junctions with topological insulator weak links can host low-energy Andreev bound states giving rise to a current-phase relation that deviates from sinusoidal behavior. Of particular interest are zero-energy Majorana bound states that form at a phase difference of $\pi$.  Here, we report on interferometry studies of Josephson junctions and superconducting quantum interference devices (SQUIDs) incorporating topological insulator weak links.  We find that the nodes in single-junction diffraction patterns and SQUID oscillations are lifted and independent of chemical potential.  At high temperatures, the SQUID oscillations revert to conventional behavior, ruling out asymmetry.  The node-lifting of the SQUID oscillations is consistent with low-energy Andreev bound states exhibiting a non-sinusoidal current-phase relation, coexisting with states possessing a conventional sinusoidal current-phase relation.  However, the finite nodal currents in the single-junction diffraction pattern suggest an anomalous contribution to the supercurrent possibly carried by Majorana bound states, though we also consider the possibility of inhomogeneity.
}
\end{abstract}

%Josephson junctions with topological insulator (TI) \cite{RevModPhys.82.3045} weak links are expected to host low-energy Andreev bound states (ABSs) giving rise to a current-phase relation (CPR) that deviates from sinusoidal behavior \cite{PhysRevLett.100.096407, PhysRevB.86.214515, PhysRevB.87.104507}.  Of particular interest are zero-energy Majorana bound states (MBS) that form either at the core of a vortex or when a phase difference of $\pi$ is imposed across the junction \cite{PhysRevLett.100.096407, Alicea2012, Beenakker2013}.  Here, we report on extensive studies of Josephson interferometry of TI junctions as a function of chemical potential and temperature.  We find that the nodes in single-junction diffraction patterns and SQUID oscillations are lifted and independent of chemical potential.  At high temperatures, the interference patterns revert to conventional behavior, ruling out the effects of asymmetry.  The node-lifting of the SQUID oscillations are consistent with low-energy ABSs exhibiting a non-sinusoidal CPR, coexisting with states possessing a conventional sinusoidal CPR. However, the finite nodal currents in the single-junction diffraction pattern suggest an anomalous contribution to the supercurrent possibly carried by MBSs \cite{PhysRevB.88.121109}.

\maketitle

\section*{Introduction}

Topological insulators (TIs) are band insulators that possess gapless, helical surface states that mimic relativistic Dirac electrons~\cite{RevModPhys.82.3045}. The helical nature of these surface states implies that with induced superconductivity one can create an analog of a $p$-wave superconductor with Majorana bound states (MBSs) at vortex cores~\cite{PhysRevLett.100.096407, Alicea2012, Beenakker2013}. By coupling a TI with two superconducting leads, one can generate a Josephson junction with anomalous current-phase relation (CPR) due to the presence of low-energy Andreev bound states (ABSs)~\cite{PhysRevLett.100.096407, PhysRevB.86.214515, PhysRevB.87.104507}. Among these states, there is a special pair with 4$\pi$ periodicity that crosses zero energy when a phase difference of $\pi$ is introduced across the junction. Such states are identified as the MBSs which obey non-Abelian exchange statistics and can be used to implement a robust topological quantum computer~\cite{RevModPhys.80.1083}.

Recently, there has been much experimental progress in realizing and studying the Josephson effect in TIs~\cite{PhysRevB.84.165120, NatCommun.2.575, NatMat.11.417, PhysRevLett.109.056803, SciRep.2.339, NatCommun.4.1689, PhysRevX.3.021007, Orlyanchik2013,  PhysRevB.89.134512, Kurter2014}.  Although most such devices appear to have a significant bulk contribution to the normal state conductance, there is evidence~\cite{NatMat.11.417, NatCommun.4.1689, PhysRevB.89.134512, Kurter2014} that the majority of the supercurrent is carried by surface states.

Josephson interferometry is a great route to study CPR of Josephson junctions. Magnetic flux threading the barrier induces phase-winding along the width of the junction, leading to interference effects that modulate the critical current. In a Josephson junction in the small junction limit with a uniform current density and a sinusoidal CPR, this results in a Fraunhofer diffraction pattern, characterized by vanishing of the critical current from destructive interference whenever an integer number of flux quanta are enclosed by the junction.  These nodes remain zero even for non-sinusoidal CPRs that are 2$\pi$-periodic.  In contrast, it has been proposed~\cite{PhysRevB.88.121109} that Josephson vortices could stabilize pairs of MBSs in TI junctions, leading to a residual critical current at integer flux quanta.  While anomalous diffraction patterns from TI junctions have been reported~\cite{PhysRevLett.109.056803}, interpretation of such node-lifting must be done carefully due to the possibility of trivial effects, such as inhomogeneity, disorder, and screening effects of large supercurrents.  Alternatively, one may analyze quantum interference between two junctions interrupting a superconducting loop to form a superconducting quantum interference device (SQUID).  In this case, flux threaded within the loop imposes a phase difference between the two junctions, generating interference that is less sensitive to junction details.

Here, we combine both approaches (single junction diffraction pattern and SQUID oscillations) to probe the CPR of TI junctions.  These approaches can be made more sensitive to unconventional supercurrent components than a direct measurement of the CPR~\cite{NanoLett.13.3086} by focusing on the nodal regimes where the sinusoidal components are canceled out by destructive interference.  We find evidence of an anomalous CPR in both the diffraction pattern and SQUID oscillations.  While a non-sinusoidal,  2$\pi$-periodic CPR can explain the node-lifting in the SQUID oscillations, this is not the case for the diffraction pattern.  Instead, we consider the possibility of MBSs, while attempting to rule out the influence of inhomogeneity in the supercurrent distribution.  Our results provide evidence for low energy ABSs in TI Josephson junctions.  While such states might include MBSs, further work is likely required to firmly exclude other effects.

\section*{Results}

\subsection*{Characterization of TI Josephson Junctions}

We analyze both single lateral Josephson junctions and dc SQUIDs incorporating them on the top surface of a thin piece of the 3D TI, Bi$_2$Se$_3$.  We focus on one particular tri-junction SQUID (illustrated in Fig.~\ref{fig:overview}a, with sample $I-V$s shown in Fig.~\ref{fig:overview}b), with similar results having been observed in many other devices.  The SQUID is formed from three superconducting leads on the surface of the TI, separated by 100 nm gaps.
  
We observe a sharp drop in the critical current with top gating (Fig.~\ref{fig:overview}c).  This signals the depletion of the conventional 2DEG originating from band-bending at the surface of Bi$_2$Se$_3$, which exposes the helical surface states that carry the majority of the supercurrent to greater disorder~\cite{Orlyanchik2013}. As carriers are depleted by the top gate we find a qualitative change in the temperature dependence (shown in Fig.~\ref{fig:overview}d), in which the junction acquires a more diffusive character~\cite{NatMat.11.417, Kurter2014}.  We have observed consistent behavior in nearly all of our TI junctions, independent of TI film thickness (from 7 to 86 nm), suggesting that the supercurrent is dominated by surface effects.  However, we emphasize that our interpretations of interferometric measurements in this paper can be made independent of exact knowledge of the role played by trivial states in the bulk or the surface.  This assertion is justified because the helical states can coexist with such trivial states~\cite{NatComm.1.128}.  Theoretical studies of doped topological superconductors~\cite{PhysRevLett.107.097001,PhysRevLett.109.237009,PhysRevB.84.144507, PhysRevB.87.035401} also suggest that the bulk can be gapped by superconductivity, permitting the observation of surface physics.

\subsection*{Diffraction Pattern and SQUID Oscillations}

In Fig.~\ref{fig:fraunhofer_and_squid}a, we show the magnetic field dependence of the critical current for two different top gate biases.  We observe rapid SQUID oscillations with a period of $\approx 0.21$ mT, consistent with the lithographic area of SQUID loop and an estimate of flux focusing by the superconducting film. The SQUID oscillations are enclosed in an envelope reflecting the diffraction pattern of the individual junctions.  Minima in this envelope correspond to integer flux quanta enclosed by the individual junctions.  We observe that the critical current does not completely vanish at these field values; instead, the current drops to a finite value which is essentially independent of gate bias. Similar node-lifting is observed in single Josephson junctions and other SQUIDs fabricated on a TI (for example, see Supplementary Figs.~1 and 2).  We also observe lifting of the nodes in the SQUID oscillations within the envelope, shown in greater detail in Fig.~\ref{fig:fraunhofer_and_squid}b.  While the maximum (antinodal) supercurrent varies dramatically with gate bias, the nodal current remains fixed at a value of roughly 150 nA near zero field. The nodal current at a fixed gate bias slowly decreases with magnetic field, much like the antinodal currents.

\subsection*{Nodal Supercurrent}

We now consider possible mechanisms for the observed node-lifting in the interference characteristics of our junctions and SQUIDs.  We focus first on the SQUID nodes because there are a number of well-known phenomena that can lift the nodes of SQUID oscillations, particularly finite inductance of the SQUID loop, parallel conductance mechanisms (that is, shorts in one of the junctions), asymmetry in junctions, and a nonsinusoidal CPR.  Because the SQUID inductance parameter $\beta = L I_{\mathrm{c}} / \Phi_0 \approx 10^{-3}$ is much less than 1 ($L$ is the loop inductance and $\Phi_0 = h / 2e$ is the magnetic flux quantum), circulating currents are unlikely to be the cause of the observed node-lifting~\cite{SQUID.Handbook}.  A superconducting short is ruled out because the node current at a given gate bias decays with field much like the anti-nodes, indicating that it is a Josephson effect spread across the junction width.

If the two junctions do not have equivalent critical currents, then perfect destructive interference will not occur at the nodes.  To test for asymmetry, we measure the nodal supercurrent at elevated temperatures and at $V_{\mathrm{TG}}=-18$ V, as illustrated in Figs.~\ref{fig:node_temp_dep}a and \ref{fig:node_temp_dep}b.  While the critical current at the first SQUID antinode (at $B = 0.146$ mT, see Fig.~\ref{fig:fraunhofer_and_squid}b) only declines gradually with temperature, we find that the nodal current at $B = 0.04$ mT collapses more rapidly and vanishes beyond 850 mK as shown in Fig.~\ref{fig:node_temp_dep}b. Part of this reduction is likely due to the suppression of the critical current by thermal fluctuations that affect the smaller nodal current more significantly.  The suppression is governed by the noise parameter $\Gamma = 2 e k_{\mathrm{B}} T / I_{\mathrm{c}}$, the ratio of the thermal energy to the Josephson coupling energy.  For $T = 800$ mK and $I_{\mathrm{c}} = 150$ nA, this gives $\Gamma = 0.22$, which should reduce the apparent critical current for an underdamped Josephson junction by a factor of $\approx 2$~\cite{PhysRevLett.22.1364}, but we observe at least a factor of 8 reduction in the nodal current upon heating. Indeed, when the SQUID antinodal critical current is suppressed to 150 nA at higher magnetic fields due to phase-winding along the junction width (for example, at $B=2.44$ mT, visible in Fig.~\ref{fig:high_temp_and_cpr}d), the detected critical current falls by 50\% between 20 mK and 800 mK, as expected. Thus, the rapid vanishing of SQUID nodal current at high temperature is inconsistent with junction asymmetry. 

\subsection*{Low Energy ABSs}

Since we have ruled out the asymmetry, an alternative explanation for the lifted SQUID nodes at lower temperature is necessary.  Indeed, our data suggests that the nodes are lifted at low temperature because of non-sinusoidal contributions to the CPR, which then become conventional or suppressed at higher temperatures.  The non-sinusoidal behavior is expected in junctions in which some of the supercurrent is carried by low energy ABSs with high transparency~\cite{PhysRevB.74.041401, PhysRevB.86.214515}.  Such unconventional CPRs are not symmetric about $\phi = \pi/2$ or $\phi = 3 \pi/2$ and do not obey $I(\phi + \pi) = -I(\phi)$; thus, when two junctions with non-sinusoidal CPRs have a $\pi$ phase difference imposed between them, as in a SQUID with flux $\Phi = \Phi_0 / 2$, the total supercurrent does not perfectly cancel out even if the two junctions are nominally identical.  In Fig.~\ref{fig:high_temp_and_cpr}a, we superimpose a simulated current-flux relation based on the formalism in Ref.~\cite{PhysRevLett.100.096407} with the observed SQUID oscillations at 20 mK, using a CPR shown in Fig.~\ref{fig:high_temp_and_cpr}b.  Here, the individual ABSs can be labeled by their transverse momentum $q$ along the width of the junction.  States with large $|q|$ contribute an essentially sinusoidal component to the CPR while states with small $|q|$ produce a highly forward-skewed component.  The two $q=0$ states are identified as the MBSs.  The nodes are lifted due to the highly forward-skewed CPR components from low $q$ (representing low energy ABSs), illustrated by the red curve in Fig.~\ref{fig:high_temp_and_cpr}b.  The blue curve in \ref{fig:high_temp_and_cpr}b represents a sinusoidal term that is added to the unconventional CPR to construct the total CPR (black trace).  We emphasize that this is a toy model to illustrate how an unconventional CPR can lead to the observed SQUID oscillations.  We have ignored details such as scattering, finite junction length, and temperature~\cite{PhysRevB.86.214515}.  A more detailed description of our model is given in the Supplementary Note 1.

In Fig.~\ref{fig:high_temp_and_cpr}c, we show the SQUID oscillations near zero field for three different temperatures.  While the SQUID nodes are prominently lifted at low temperature, beyond 800 mK the nodes are fully formed and the SQUID oscillations better described by two essentially identical junctions with sinusoidal CPR, suggesting that the anomalous components revert to a conventional sinusoidal form.  At higher temperature the CPR becomes conventional, consistent with direct measurements of CPR in SNS devices~\cite{RevModPhys.76.411} and expected in the case of TI Josephson junctions~\cite{PhysRevB.86.214515}.  Our direct measurements of the CPR in TI junctions also show evidence of slightly forward skewness that disappears with temperature (see Supplementary Fig.~3 and Supplementary Methods).  The independence of the SQUID nodal supercurrent with gating (shown in Fig.~\ref{fig:fraunhofer_and_squid}b) suggests that the top gate primarily suppresses conventional states.

One might claim that the additional current at the nodes comes from some separate component with a sinusoidal CPR, such as current through the bulk or the bottom layer.  This hypothetical component could conceivably be asymmetric between the two junctions, not affected by the top gate, and much more susceptible to increased temperature (due to lower mobility or phase coherence).  However, such a component should also contribute a sizable amount to the antinode current.  But at $V_{\mathrm{TG}}=-18$ V, the antinode current is largely temperature independent from base temperature up to 800 mK.

\section*{Discussion}

Having established that non-sinusoidal terms in the CPR arising from low energy ABSs can lift the SQUID nodes, we return to the diffraction pattern envelope to consider the origin of the lifting at the nodes of the single junction diffraction pattern.  It is important to note that skewed CPR components from $q \neq 0$ modes can lift the nodes of the SQUID oscillations, but these would still undergo completely destructive interference at the nodes of the diffraction pattern envelope as long as the CPR is $2\pi$ periodic.  It is well-known that distortions from an ideal Fraunhofer diffraction pattern can arise from inhomogeneous critical current distribution and can lead to lifted nodes in the diffraction pattern due to an incomplete cancellation of supercurrent.  This effect is difficult to rule out.  However, we do not believe that critical current inhomogeneity can fully explain the lifting of the nodes in the SQUID oscillations and single junction diffraction pattern for several reasons.  First, the small amount of asymmetry between the two junctions in the SQUID strongly suggests that such inhomogeneity is low (see Supplementary Fig.~4 and Supplementary Note 2 for a comparison between a simple model of inhomogeneity and the observed lifting of the first node in the diffraction pattern envelope).  In addition, the diffraction patterns we observe in nearly all of our junctions exhibit a strongly lifted first node and a nearly vanishing second node.  This can be seen in the 800 mK trace in Fig.~\ref{fig:high_temp_and_cpr}d (where the second node in the diffraction pattern envelope is lower than both the first and third node) and in single junction diffraction patterns (see Supplementary Fig.~1 and Ref.~\cite{Orlyanchik2013}).  This shape is difficult to achieve solely by critical current asymmetry and its consistency across many junctions makes us consider alternative explanations for the single junction node lifting.

It has been proposed that lifted notes in the diffraction pattern can arise from the hybridization of pairs of MBSs that exist on the top and bottom surfaces of the topological insulator barrier~\cite{PhysRevB.88.121109}.  These are bound to Josephson vortices at the locations where the phase difference is an odd multiple of $\pi$.  When there is an integer multiple of flux quanta in the junction, at which a node is expected, these states move to the edge of the junction and the MBSs hybridize, creating a burst of supercurrent and lifting the nodes by effectively generating a strong asymmetry in junction critical current distribution. The magnitude of the observed nodal supercurrent agrees with the predicted value of $I \approx \Delta_0 / \Phi_0 \approx 100$ nA, where $\Delta_0$ is the niobium superconducting gap and $\Phi_0 = h/2e$ is the magnetic flux quantum.  Our use of the model in Ref.~\cite{PhysRevB.88.121109} is justifiable because the thickness of our TI films (6 to 30 nm) is smaller than the spatial size of the MBSs (as large as tens of nanometers), allowing us to ignore additional ABSs at the edge of the junction.  The independence of the diffraction pattern nodal current with respect to top gate bias is also consistent with the robustness of MBSs to changes in chemical potential.

We admit that some features of our data are not readily described by this model invoking MBSs.  For example, this mechanism involves low energy ABSs that would also contribute to the lifting of the SQUID nodes.  However, we find in Fig.~\ref{fig:high_temp_and_cpr}d that the diffraction pattern envelope nodes do not collapse at 800 mK, where the SQUID nodes vanish.  More significantly, this model predicts that all junction nodes are lifted equally, which conflicts with our observation that the odd nodes tend to be lifted preferentially to the even ones.  We speculate that hybridization of MBSs in adjacent vortices might alter the details of the node-lifting and distinguish between an even and odd number of MBSs pairs.  One can understand this by noting that while an even number of MBS pairs can fully hybridize each other and remove any zero energy states, for an odd number there will always be at least a single pair remaining.  Nonetheless, these discrepancies as well as the possibility of inhomogeneity force us to be cautious about identifying the anomalous diffraction patterns as firm evidence for MBSs.

%We note that a $\sin{(\phi/2)}$ component in the CPR would also lead naturally to the observed even-odd effect in the lifting of the nodes in the diffraction pattern envelope and contribute to a uniform lifting of the SQUID nodes.  This could arise from the MBSs that form zero energy states in the junction wherever the phase difference is an odd multiple of $\pi$, as in Ref.~\cite{PhysRevB.88.121109}.  While parity transitions might suppress such a contribution or convert it to a conventional 2$\pi$ periodicity, our measurements occur at finite voltage once we exceed the critical current. Thus, the Josephson phase evolution should become rapid enough (easily in the GHz range for $\mu$V voltage differences) to allow Zener tunneling through the small gap caused by parity transitions (typically in the kilohertz range) and reveal the influence of 4$\pi$ periodic states on the current-voltage characteristics of the junction \cite{PhysRevB.86.024509, PhysRevB.87.104507, arxiv.1403.2747}.  A full dynamical calculation of the phase dynamics and parity transition rates will be required to determine that.

\section*{Methods}

\subsection*{Sample preparation}
Single crystals of Bi$_2$Se$_3$ were grown by melting a mixture of pure Bi and Se in a stoichiometric ratio of 1.9975:3 (Bi:Se) in a vacuum quartz tube at 800 $^{\circ}$C.  Thin flakes (6-30 nm) of Bi$_2$Se$_3$ were exfoliated onto silicon substrates covered by a 300 nm thick SiO$_2$ layer.  The sample in the main text is 9 nm thick.  Such thin flakes typically have a 2D carrier density of $N_{\mathrm{2D}} \sim10^{13} - 10^{14}$ cm$^{-2}$ and low temperature mobility $\mu \sim 10^2 - 10^3$ cm$^{2}$/V-s, as determined by Hall bar measurements on separate flakes of similar dimensions.  Weak antilocalization measurements of such Hall bars give typical phase-coherence lengths of $\ell_{\phi}= 300$ - 1000 nm at 10 mK.  Superconducting leads were defined by conventional e-beam lithography and a subsequent DC sputtering of 50 nm of Nb at room temperature.  Brief Ar ion milling is employed before metallization \emph{in situ} to ensure good contact between the Bi$_2$Se$_3$ and the leads.  A top gate may be created by covering the sample with 30 nm of alumina via ALD and deposition of Ti/Au over the exposed Bi$_2$Se$_3$.

\subsection*{Low temperature measurement}

The devices were thermally anchored to the mixing chamber of a cryogen-free dilution refrigerator equipped with a vector magnet and filtered wiring.  We perform current-biased transport measurements with standard lockin techniques, typically with a 4 nA AC excitation at $f=73$ Hz.  The doped silicon substrate can act as an electrostatic back gate, but we found that the critical current was only very weakly tuned by back gate bias.  The device featured in the main part of this paper was 9 nm thick and possessed a normal state resistance of 37 ohms, which was only weakly dependent on top or back gate bias.  All data in the main section of this paper were taken at 20 mK, unless stated otherwise.  We plot and report all data in terms of applied magnetic field.  A small amount of magnetic field ($B < 0.2$ mT) is present even at zero applied field, likely due to magnetic flux trapped within the superconducting magnet.  To achieve zero effective field, we tune the applied field until the supercurrent is maximized (that is, there is no destructive interference from residual fields).

\section*{Acknowledgements}

We acknowledge helpful discussions with Liang Fu, Pouyan Ghaemi, Taylor Hughes, Christopher Nugroho, Vladimir Orlyanchik, and Martin Stehno.  C.K., A.D.K.F., and D.J.V.H.~acknowledge funding by Microsoft Project Q.  Y.S.H.~acknowledges support from National Science Foundation grant DMR-12-55607.  Device fabrication was carried out in the MRL Central Facilities (partially supported by the DOE under DE-FG02-07ER46453 and DE-FG02-07ER46471).

\section*{Contributions}
Y.S.H.~grew the bismuth selenide crystals.  C.K.~and A.D.K.F.~fabricated the devices and performed the measurements.  All authors analyzed the data and wrote the manuscript.

\section*{Competing Financial Interests}
The authors report no competing financial interests.

\bibliography{dcsquidbib}

\newpage

\dograph{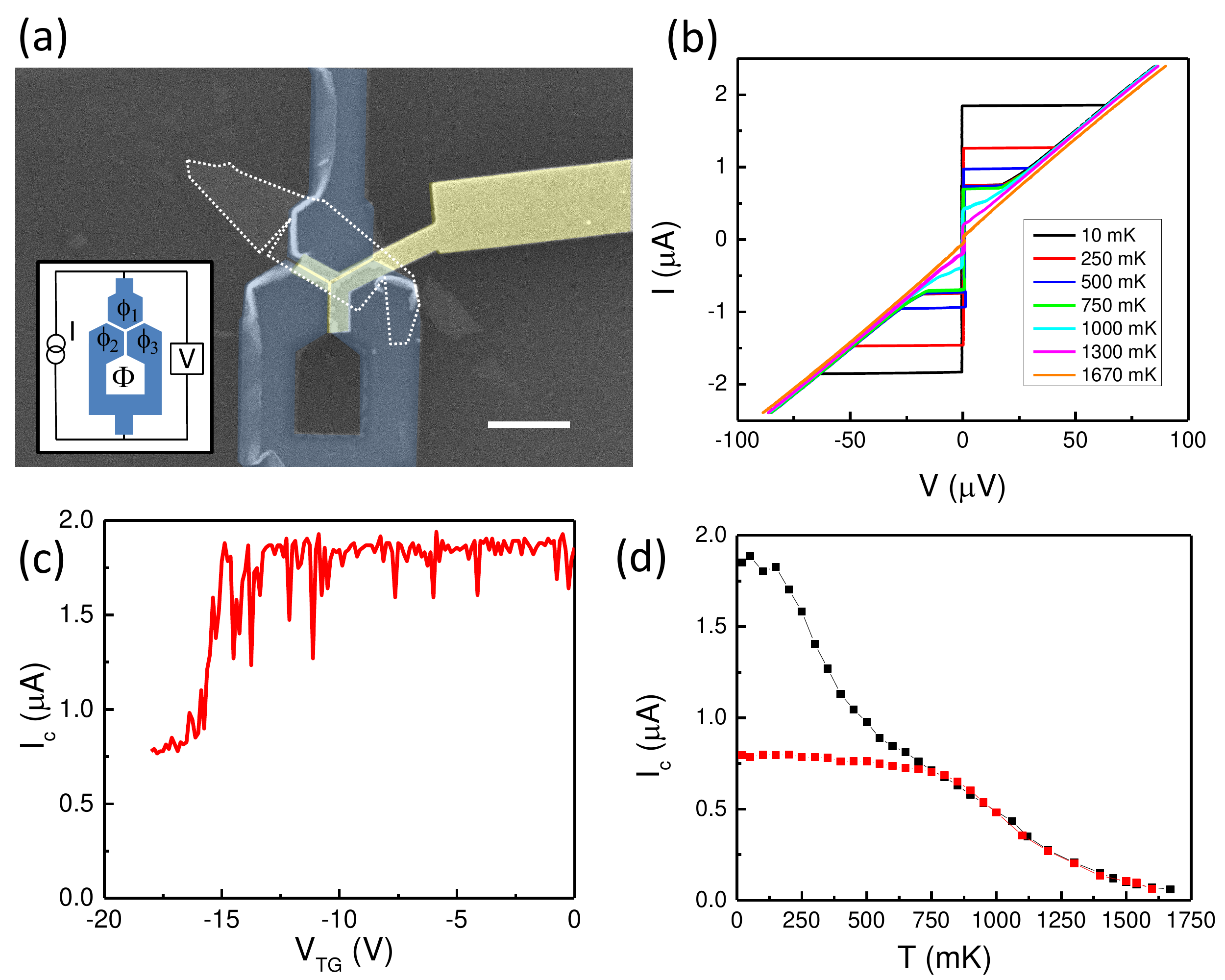}{{\bf Characterization of SQUID} (a) SEM image of device.  The Bi$_2$Se$_3$ flake is outlined with a dashed white line.  Top gate is colored yellow and the niobium leads are colored blue.  Each junction has a length of $\sim 100$ nm and a width of 1 micron.  Scale bar is 2 microns.  Schematic of device (TI flake and top gate not shown) is shown in inset.  Our measurements are sensitive to the sum of the critical currents of two of the junctions, with a phase difference $\phi_2 - \phi_3 = 2\pi(\Phi)/(\Phi_0)$ set by the magnetic flux $\Phi$ within the superconducting loop.  Here, $\Phi_0$ is the magnetic flux quantum.  The third junction is not directly probed here.  It only partially covers the TI flake and thus only weakly modifies the much larger supercurrent circulating around the loop.  (b) $I-V$s vs temperature at $B=0$ and $V_{\mathrm{TG}}=0$ V, clearly demonstrating zero resistance state. (c) Top gate dependence, showing an abrupt drop in supercurrent.  (d) Temperature dependence of critical current for $V_{\mathrm{TG}}=0$ (black squares) and $V_{\mathrm{TG}}=-18$ V (red squares), showing a transformation from ballistic to diffusive behavior.  Note that the supercurrent at low density ($V_{\mathrm{TG}}=-18$ V) is only weakly dependent on temperature up to 800 mK.}

\dograph{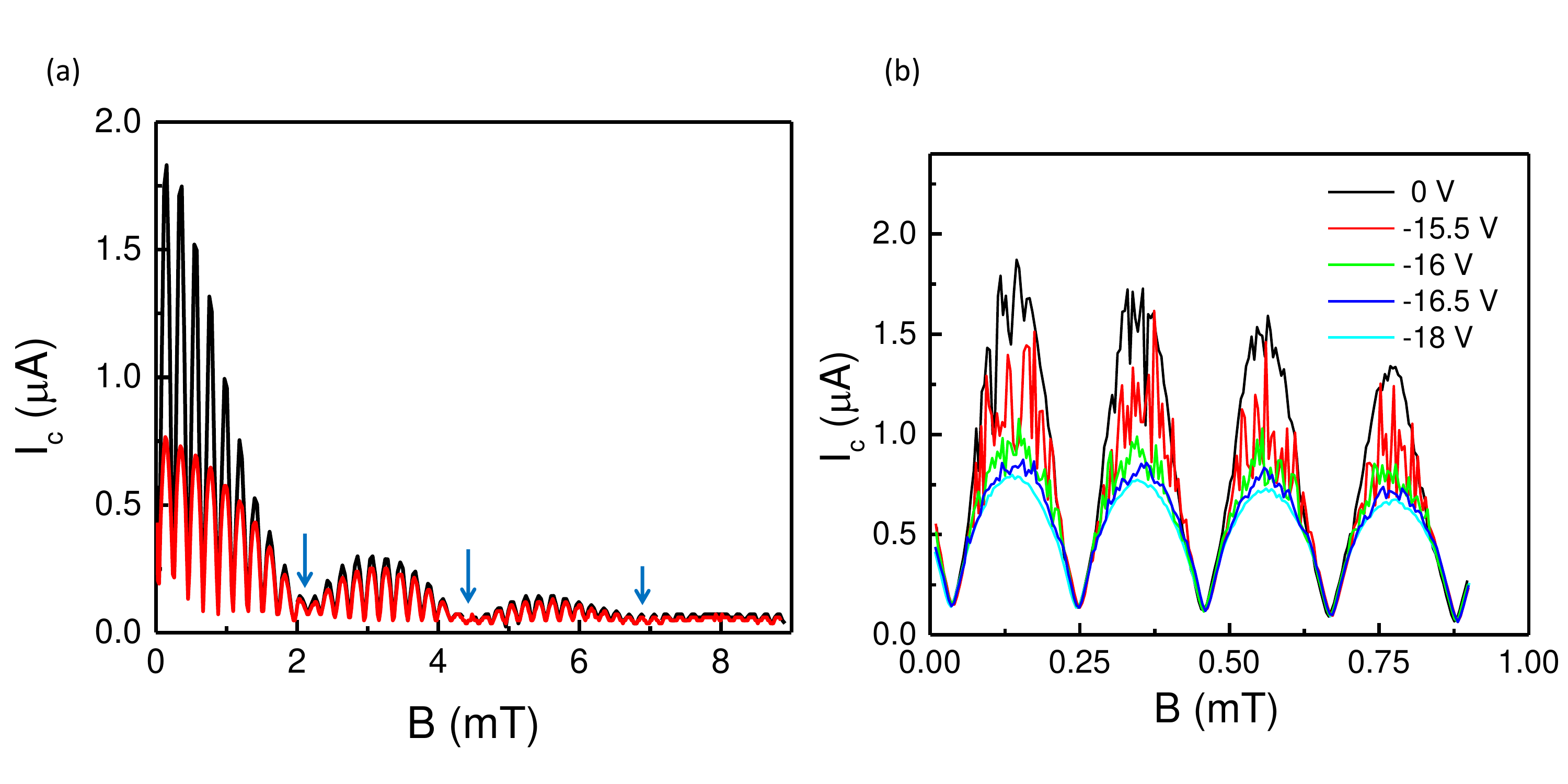}{{\bf Phase-sensitive properties of supercurrent} (a) $I_{\mathrm{c}}$ vs $B$ traces for $V_{\mathrm{TG}}=0$ (black) and $V_{\mathrm{TG}}=-18$ V (red) at 20 mK, showing both high frequency SQUID oscillations bound by an envelope corresponding to the single-junction diffraction patterns.  Blue arrows point to nodes in diffraction pattern. (b) Additional data sets with high magnetic field resolution to show SQUID oscillations at 20 mK near zero field for various gate values.  Note the gate independence of nodes in both (a) and (b), though the SQUID nodes slowly decrease with magnetic field.}

\dograph{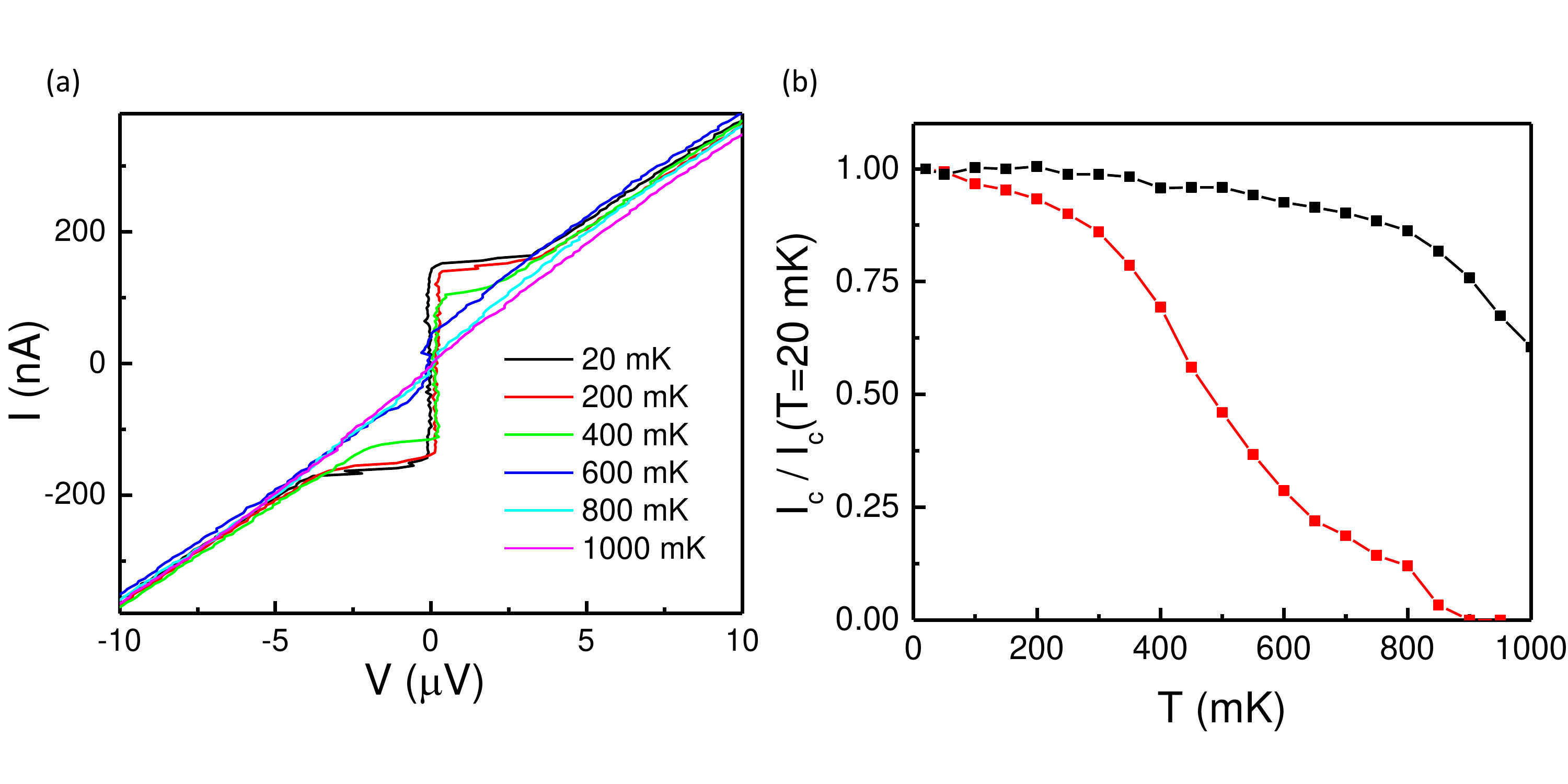}{{\bf Suppression of SQUID nodal current at elevated temperature} (a) Current-voltage curves at the SQUID node closest to zero applied field ($B = 0.04$ mT) and $V_{\mathrm{TG}} = -18$ V.   (b)  Temperature dependence of SQUID nodal ($B = 0.04$ mT, black squares) and SQUID antinodal ($B=0.146$ mT, red squares) critical currents; both sets are normalized by their values at 20 mK and were taken at $V_{\mathrm{TG}} = -18$ V.  In both cases, current-voltage curves were taken as close to zero applied field as possible, where there is a minimal amount of magnetic flux threading the individual junctions.  For rounded current-voltage curves, we use the "excess current" method to determine critical current.}

\dograph{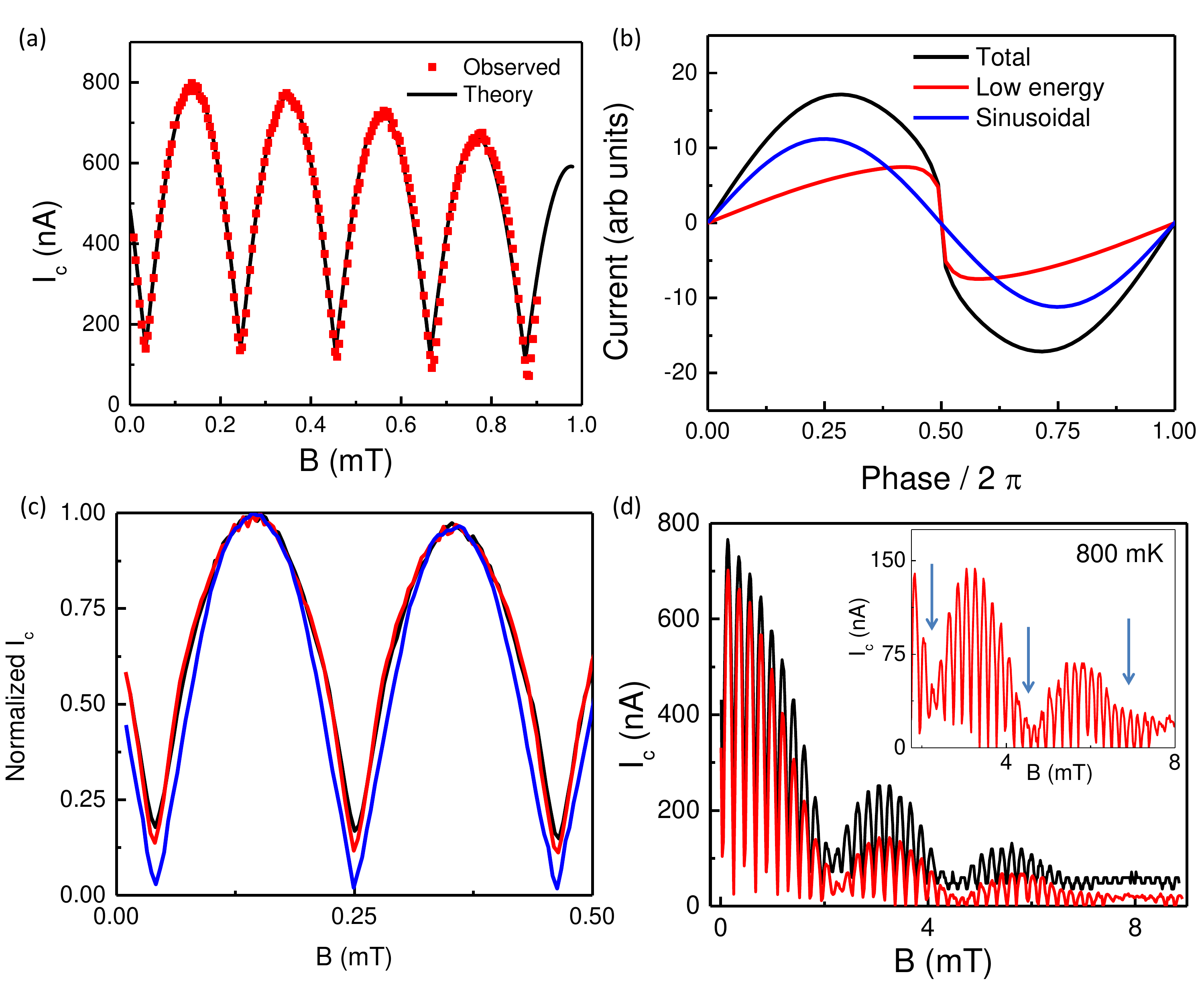}{{\bf SQUID oscillation model and high temperature behavior} (a) Observed (red squares) SQUID oscillations for $V_{\mathrm{TG}}=-18$ V and 20 mK, compared with theoretical (black) curve based on a toy model CPR.  (b) Theoretical CPR used to derive the theoretical SQUID oscillations in (a).  We add the $q=0$ and $q=0.1$ modes (in normalized units in which the velocity $v$ and energy gap $\Delta_0$ are set equal to 1) to a purely sinusoidal CPR (shown in blue).  The resulting composite CPR is shown in black, with the contribution from the $q=0$ and $q=0.1$ modes shown in red.  Note that the subtle non-sinusoidal behavior leads to detectable levels of node-lifting in SQUID oscillations while possibly being obscured in direct measurement of CPR.  The theoretical SQUID oscillations were rescaled by $\sin{(\pi B / B_0)}/(\pi B / B_0)$ to mimic Fraunhofer-like decay of critical current due to single-junction diffraction.  (c) Critical current vs magnetic field for 20 mK (black), 400 mK (red), and 800 mK (blue) at $V_{\mathrm{TG}} = -18$ V, showing a discernible change in SQUID modulation depth.  For each temperature, the critical current is normalized by the value at $B=0.146$ mT.  As the temperature increases, the current at the SQUID nodes decreases indicating that the CPR is reverting to a conventional form (that is, sinusoidal).  (d)  Comparison of $I_{\mathrm{c}}$ vs $B$ traces at $V_{\mathrm{TG}} = -18$ V for 20 mK (black) and 800 mK (red).  Inset shows the same data as the 800 mK in detail to show the diffraction pattern nodes (blue arrows).}

\end{document}